\documentstyle[twocolumn,aps,epsf,array]{revtex}
\begin{document}
\draft
\title{Kochen-Specker theorem and \\
experimental test on hidden
variables\thanks{To appear in {\em Int. J. Mod. Phys. A.}}}
\author{Ad\'{a}n Cabello\thanks{Electronic address:
fite1z1@sis.ucm.es}}
\address{Departamento de F\'{\i}sica Aplicada,
Universidad de Sevilla, 41012 Sevilla, Spain}
\date{\today}
\maketitle
\begin{abstract}
A recent proposal to experimentally test
quantum mechanics against noncontextual hidden-variable
theories [{\em Phys. Rev. Lett.} {\bf 80}, 1797 (1998)]
is shown to be related with
the smallest proof of the Kochen-Specker
theorem currently known
[{\em Phys. Lett. A} {\bf 212}, 183 (1996)].
This proof contains eighteen yes-no questions about a
four-dimensional physical system,
combined in nine mutually incompatible tests.
When these tests are
considered as tests about a two-part two-state system,
then quantum mechanics and non-contextual hidden variables
make the same predictions for eight of them,
but make different predictions for the ninth.
Therefore, this ninth test would allow us to
discriminate between quantum mechanics and
noncontextual hidden-variable
theories in a ({\em gedanken}) single run experiment.
\end{abstract}

\narrowtext

\section{The Kochen-Specker theorem}
The Kochen-Specker (KS) theorem \cite{Specker60,Bell66,KS67}
contains one of the most fundamental findings in
quantum mechanics (QM):
Yes-no questions about an individual physical system cannot be
assigned a unique answer in such a way that the result of measuring
any mutually commuting subset of these yes-no questions can be
interpreted as revealing these preexisting answers.
To be precise, the KS theorem asserts that,
in a Hilbert space with a finite dimension,
$d\ge 3$, it is possible to construct a set of $n$ projection operators,
which represent yes-no questions about an individual physical system,
so that none of the $2^n$ possible sets of ``yes'' or ``no'' answers
is compatible with the sum rule of QM for orthogonal resolutions of
the identity (i.e., if the sum of a subset of mutually orthogonal
projection operators is the identity, one and only one of the
corresponding answers ought to be ``yes'').
This conclusion holds irrespective of the
quantum state of the system.
Implicit in the KS theorem is the assumption of
{\em noncontextuality}: Each yes-no question is assigned a single
unique answer, independent of which subset of mutually commuting
projection operators one might consider it with.
Therefore, the KS theorem discards
hidden-variable theories with this property,
known as noncontextual
hidden-variable (NCHV) theories. Local hidden-variable theories,
such as those discarded by
Bell's theorem \cite{Bell64}, are a particular type of NCHV theories, so
in this sense, the KS theorem is more general than Bell's theorem.
However, while Bell's theorem has been
successfully tested in the laboratory \cite{lab},
the translation of
any proof of the KS theorem into
an experiment on an individual system seems to be an impossible task.
This is so because the $n$ projection operators
appearing in a proof of the KS theorem can be combined in
$t$ different orthogonal resolutions of
the identity, and each of them represents a maximal test
which is incompatible with the other $t-1$ maximal tests.
However, we have recently presented an experiment which
seems to challenge the idea that the KS theorem
cannot be tested in the laboratory: A test on an individual
system of a two-part two-state system (prepared in whatever
quantum state) exists for which the predictions of
any NCHV theory always differ with those of QM \cite{CG98}.

This test would be related somehow with some of the
proofs of the KS theorem.
The smallest proof of the KS theorem
currently known contains eighteen
projection operators (yes-no questions) which can be combined in nine
resolutions of the identity (maximal tests) of a four-dimensional
Hilbert space, ${\cal H}_4$ \cite{CEG96,simple}.
Both results, \cite{CG98} and \cite{CEG96}, concern
to physical systems described in ${\cal H}_4$, and
it would be interesting to establish which is the connection
between them \cite{Lucien}.
That is precisely the aim of this paper.

\section{The proof with 18 vectors}
The proof of the KS theorem with eighteen
projection operators
in ${\cal H}_4$ is given in Table I \cite{Peres95}.

\begin{center}
\begin{tabular}{|ccccccccc|}
\hline
\hline
$1000$ & $1111$ & $1111$ & $1000$ &
$1001$ & $1001$ & $111\bar{1}$ & $111\bar{1}$ & $100\bar{1}$ \\
$0100$ & $11\bar{1}\bar{1}$ & $1\bar{1}1\bar{1}$ & $0010$ &
$0100$ & $1\bar{1}1\bar{1}$ & $1\bar{1}00$ & $0101$ & $0110$ \\
$0011$ & $1\bar{1}00$ & $10\bar{1}0$ & $0101$ &
$0010$ & $11\bar{1}\bar{1}$ & $0011$ & $10\bar{1}0$ & $11\bar{1}1$ \\
$001\bar{1}$ & $001\bar{1}$ & $010\bar{1}$ & $010\bar{1}$ &
$100\bar{1}$ & $0110$ & $11\bar{1}1$ & $1\bar{1}11$ & $1\bar{1}11$ \\
\hline
\hline
\end{tabular}
\end{center}
\begin{center}
\noindent TABLE I:
{\small Proof of the Kochen-Specker theorem in ${\cal H}_{4}$.\\}
\end{center}

\noindent Table I contains eighteen vectors combined in nine columns.
Each vector represents the projection operator
onto the corresponding normalized vector.
For instance, $001\bar{1}$
represents the projector onto the vector
${\scriptstyle \frac{1}{\sqrt 2}}(0,0,1,-1)$.
Each column contains four mutually orthogonal vectors,
so that the corresponding projectors sum the identity in ${\cal H}_4$.
Therefore,
in a NCHV theory, each column must have assigned the answer
``yes'' to one and only one vector.
But it is easily seen that such an assignment
is impossible since each vector in Table I appears twice,
so that the total number of ``yes'' answers must be an even number.

If we examine this proof without noticing
which particular physical system it refers to, all we
see is that each projector in Table I is orthogonal to other seven
projectors and belongs to two distinct resolutions of the identity.
In this sense, all involved projectors and resolutions
of the identity play the same role in the proof.
This absence of privileged yes-no questions or
tests is also characteristic
of every proof of the KS theorem in ${\cal H}_3$ \cite{Peres93}.

However, in ${\cal H}_4$ this situation of apparent symmetry changes
if ${\cal H}_4$ can be viewed as a product of two tensor
factors, ${\cal H}_2 \otimes {\cal H}_2$, corresponding
to some subsystems of the physical system.
Examples of systems
in which such a decomposition is physically
meaningful
(in a sense that will be specified in Sec. III)
are two-part two-level
systems such as
two spin-$\frac{1}{2}$
particles without translational motion, the polarization state of
two photons, or two internal levels of a pair of trapped ions.
Examples of systems in which such a decomposition is
mathematically possible
but will not have the same physical meaning
mentioned above are
the spin state of a single spin-$\frac{3}{2}$ particle or
two different two-level degrees of freedom in a single ion.

In the following I will suppose that ${\cal H}_4$ represents
the spin state of two spin-$\frac{1}{2}$
particles. Then the translation from the proof in Table I into
a proof in
${\cal H}_2 \otimes {\cal H}_2$ can be easily
achieved by realizing that the eighteen vectors in Table I
are eigenvectors
of some products of the usual
representation of the Pauli matrices
$\sigma_{z}$ and $\sigma_{x}$ for the spin state of spin-$\frac{1}{2}$
particles. Then Table I can be rewritten as Table II.

\begin{center}
\begin{tabular}{|cccc|cccc|c|}
\hline
\hline
$\,\,zz\,\,$ & $\,\,xx\,\,$ & $\,\,xx\,\,$ & $\,\,zz\,\,$ &
$zzxx$ & $zzxx$ & $zxxz$ & $zxxz$ & $zz\overline{xx}$ \\
$z\bar{z}$ & $\bar{x}x$ & $x\bar{x}$ &
$\bar{z}z$ &
$z\bar{z}$ & $x\bar{x}$ & $z\bar{x}$ &
$x\bar{z}$ & $\overline{zz}xx$ \\
$\bar{z}x$ & $z\bar{x}$ & $\bar{x}z$ &
$x\bar{z}$ &
$\bar{z}z$ & $\bar{x}x$ & $\bar{z}x$ &
$\bar{x}z$ & $zx\overline{xz}$ \\
$\bar{z}\bar{x}$ & $\bar{z}\bar{x}$ & $\bar{x}\bar{z}$ &
$\bar{x}\bar{z}$ &
$zz\overline{xx}$ & $\overline{zz}xx$ & $zx\overline{xz}$ &
$\overline{zx}xz$ & $\overline{zx}xz$ \\
\hline
\hline
\end{tabular}
\end{center}
\begin{center}
\noindent TABLE II:
{\small Proof of the Kochen-Specker theorem
in ${\cal H}_{2} \otimes {\cal H}_{2}$.\\}
\end{center}

\noindent The notation of Table II is the following:
$z\bar{x}$ represents the yes-no question ``are the spin component
of first particle positive
in the $z$ direction and the spin component of second particle
negative in the $x$ direction?'', and
$\overline{zx}xz$ denotes the yes-no question
``are the products
$zx:=\sigma _{1z}\otimes\sigma _{2x}$ and
$xz:=\sigma _{1x}\otimes\sigma _{2z}$ negative and
positive respectively?'', etc.
The first is an example of a
{\em factorizable} yes-no question, since it can be answered
after separate tests on the first and second particles.
The latter is an example of an
{\em entangled} yes-no question, since it
cannot be answered after separate tests on both particles.
Therefore, in Table II there are two types of yes-no questions
and, consequently, three types of maximal tests:
those involving
factorizable yes-no questions only,
such as those in columns 1 to 4;
those involving both factorizable and
entangled yes-no questions, such as those in columns 4 to 8;
and those involving entangled yes-no questions only, such as
the one in the ninth column. Taking into account this
new hierarchy of experiments, the relevant elements of
the proof of the KS theorem in ${\cal H}_2 \otimes {\cal H}_2$
can be illustrated as in Fig. 1.

\begin{figure}
\epsfxsize=8cm
\epsfbox{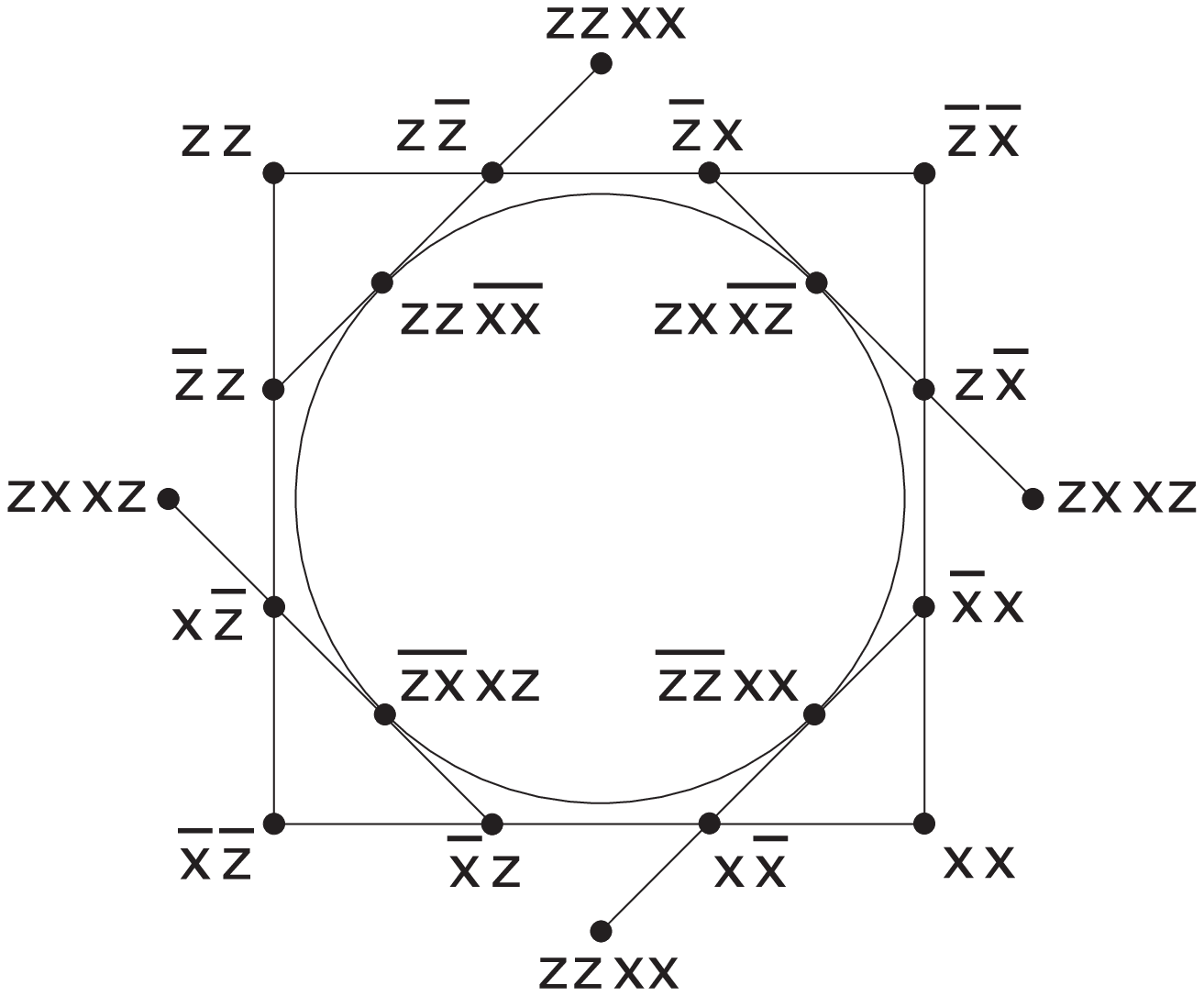}
\end{figure}
\noindent FIG. 1:
{\small Hierarchy of tests in
the proof of the Kochen-Specker theorem
in ${\cal H}_{2} \otimes {\cal H}_{2}$.
Each dot represents a yes-no question.
The upper and the lower
dots both represent the same yes-no question,
and the far left and the far right dots
both represent another yes-no question.
Dots in the same straight line or in the same circumference
represent mutually compatible yes-no questions,
and therefore each
straight line and circumference represents a
maximal test. Each straight line belonging to the square
represents a test containing only factorizable
yes-no questions. The other straight lines
represent tests containing both factorizable
and entangled yes-no questions.
The circumference represents a test
containing only entangled yes-no questions.\\}

In the next section I will show that the algebraic contradiction
contained in this proof of the KS theorem
in ${\cal H}_2 \otimes {\cal H}_2$ is
of a different kind to that of those appearing in the proofs
of the KS theorem in ${\cal H}_3$.
While in ${\cal H}_3$ there are no conflicting predictions
between QM and NCHV theories for any of the single tests
appearing in the proofs (basically
because NCHV theories do not make specific predictions
apart from those of QM), in ${\cal H}_2 \otimes {\cal H}_2$
NCHV theories can make specific predictions for single tests
that may disagree with those of QM.

\section{Noncontextual hidden-variable
theories}
A NCHV theory must satisfy the following assumptions:

(i) Any {\em one-particle observable} must have a definite value.
This is not in contradiction with QM since the Kochen-Specker
theorem is not valid for ${\cal H}_2$. In fact, in ${\cal H}_2$
specific NCHV models exist reproducing all statistical
predictions of QM \cite{NCHV}.
This is one of the reasons why the decomposition
of ${\cal H}_4$ into ${\cal H}_2 \otimes {\cal H}_2$
makes more physical sense in some systems than in others.
For our purposes we will assume that the observables
$z_1:=\sigma _{1z}$,
$x_1:=\sigma _{1x}$, $z_2:=\sigma _{2z}$,
and $x_2:=\sigma _{2x}$ must
have predefined values, either $+1$ or $-1$
(that will be denoted simply by ``$+$'' or ``$-$'').
Therefore, there are 16 possible distinct states (in a NCHV
theory), corresponding to the different
combinations of possible values for these four one-particle
observables.

(ii) The value of a {\em two-particle observable}
which is a product of two one-particle observables
corresponding to different particles, such as
$z_{1}z_{2}:=\sigma _{1z}\otimes\sigma _{2z}$,
$x_{1}x_{2}:=\sigma _{1x}\otimes\sigma _{2x}$,
$z_{1}x_{2}:=\sigma _{1z}\otimes\sigma _{2x}$, and
$x_{1}z_{2}:=\sigma _{1x}\otimes\sigma _{2z}$,
is the product of values of the corresponding one-particle
observables.
This assumption is a consequence of
the general assumption of noncontextuality.
Consider, for instance, the obsevable $zz$.
One particular way of measuring $zz$ is by measuring
$z_1$ and $z_2$ separately and multiplying
their results. The independece of
the results of the separated measurements is garanteed
(if such measurements are spacelike separated)
by the highest form of noncontextuality: locality
(so this the second reason why the decomposition
of ${\cal H}_4$ into ${\cal H}_2 \otimes {\cal H}_2$
makes more physical sense in some systems than in others).
Then the definition of noncontextuality
entails that the value for $zz$ must be the same whatever
the experimental context one might choose \cite{remark}.

(iii) The answer to a {\em yes-no question}
is logically related with the values of the involved observables.
For instance, the answer to the question $zz\overline{xx}$
is ``yes'' if $z_{1} = z_{2}$ and $x_{1} = -x_{2}$, and
``no'' in any other case.

Let us now examine the predictions of a NCHV theory
for the nine tests. Table III contains all possible
values for one-particle and two-particle observables involved
in the proof.

\begin{center}
\begin{tabular}{|cccc|cccc|cccc|}
\hline
\hline
\multicolumn{4}{|c|}{one-particle} &
\multicolumn{4}{c|}{two-particle} &
\multicolumn{4}{c|}{two-particle} \\
\multicolumn{4}{|c|}{observables} &
\multicolumn{4}{c|}{observables} &
\multicolumn{4}{c|}{yes-no questions} \\
\hline
$z_1$ & $x_1$ & $z_2$ & $x_2$ &
$z_{1}z_{2}$ & $x_{1}x_{2}$ & $z_{1}x_{2}$ & $x_{1}z_{2}$ &
$zz\overline{xx}$ & $\overline{zz}xx$ & $zx\overline{xz}$ &
$\overline{zx}xz$ \\
\hline
$\pm$ & $\pm$ & $\pm$ & $\pm$ &
$+$ & $+$ & $+$ & $+$ &
no & no & no & no \\
$\pm$ & $\pm$ & $\pm$ & $\mp$ &
$+$ & $-$ & $-$ & $+$ &
yes & no & no & yes \\
$\pm$ & $\pm$ & $\mp$ & $\pm$ &
$-$ & $+$ & $+$ & $-$ &
no & yes & yes & no \\
$\pm$ & $\pm$ & $\mp$ & $\mp$ &
$-$ & $-$ & $-$ & $-$ &
no & no & no & no \\
$\pm$ & $\mp$ & $\pm$ & $\pm$ &
$+$ & $-$ & $+$ & $-$ &
yes & no & yes & no \\
$\pm$ & $\mp$ & $\pm$ & $\mp$ &
$+$ & $+$ & $-$ & $-$ &
no & no & no & no \\
$\pm$ & $\mp$ & $\mp$ & $\pm$ &
$-$ & $-$ & $+$ & $+$ &
no & no & no & no \\
$\pm$ & $\mp$ & $\mp$ & $\mp$ &
$-$ & $+$ & $-$ & $+$ &
no & yes & no & yes \\
\hline
\hline
\end{tabular}
\end{center}
\begin{center}
\noindent TABLE III:
{\small Possible values in a NCHV theory
for the observables and yes-no questions involved in the
ninth test of the KS theorem in ${\cal H}_2 \otimes {\cal H}_2$.\\}
\end{center}

\noindent As can easily be seen studying Table III:

(a) For each of the four tests
involving only factorizable yes-no questions (columns 1 to 4 in
Table II) NCHV theories predict that
one and only one of the answers must be ``yes''.
The same prediction as in QM.
Note that if, instead of $z$ and $x$, we
choose any other spin component, then the predictions
of both QM and NCHV theories will still agree.

(b) For each of the four tests involving
factorizable and entangled
yes-no questions (columns 5 to 8 in Table II)
NCHV theories also predict that
one and only one of the answers must be ``yes''.
As QM does.
Note that the choice of spin
components $z$ and $x$ allows the corresponding
two-particle observables to have a common
eigenvalue so the corresponding yes-no question
can be represented in QM by a projection operator.

(c) Finally, consider the last test, involving
only entangled yes-no questions (column 9 in Table II).
If one checks Table III,
one reaches the conclusion that in a NCHV theory
the four yes-no questions appearing in column 9 in Table II
are {\em not} mutually {\em exclusive} (since the answers to
two of them can be ``yes'') nor {\em not exhaustive}
(since the answers to all of them can be ``no'').
The quantum prediction is radically different.
In QM, these four yes-no questions
(or, to be precise, their corresponding projectors)
form an orthogonal resolution of the identity,
so they represent mutually exclusive and exhaustive questions:
the answers must be one ``yes'' and three ``no''.

\section{Conclusion and further developments}
In brief, the predictions of QM and NCHV agree
for the first eight tests of the eighteen vector's proof of the KS theorem
in ${\cal H}_2 \otimes {\cal H}_2$, but disagree for the ninth test.
This clarifies the relationship between this proof, originally proposed in
\cite{CEG96}, and the test on NCHV theories proposed in \cite{CG98}.

A joint measurement
on a single system of the
four projection operators appearing in the ninth test
is closely related \cite{CG98} to the problem
of performing a measurement of a nondegenerate
Bell operator \cite{bellop}. Such a measurement is also
required for reliable (i.e., with 100\% theoretical
probability of success) double density quantum coding \cite{qdc1} and
for reliable teleportation \cite{telep1}.
It has recently been proved \cite{Vaidman,LCS}
that a measurement of a nondegenerate
Bell operator cannot be
achieved without using quantum systems interacting one
with the other, a condition which is not fulfilled
in recent experiments \cite{qdc2,telep2},
but which could be achieved with atoms and
electro-magnetic cavities \cite{Vaidman}
in near-future experiments.\\

\section*{Aknowledgments}
This paper is the response to a question raised by Lucien Hardy after
a talk at Oxford University in May 1998. I would like to thank
Mark S. Child and Gon\-za\-lo Gar\-c\'{\i}a de Po\-la\-vie\-ja for
inviting me to give that talk. I also acknowledge useful conversations on
this subject with Ignacio Cirac, David Mermin, Guillermo Garc\'{\i}a
Alcaine, Asher Peres, and Emilio Santos, and stimulating comments by
Helle Bechmann Pasquinucci, Sibasish Ghosh, Gebhard Gr\"{u}bl,
and R. L. Schafir.
This work was financially supported by the Universidad de Sevilla
(OGICYT-191-97) and the Junta de Andaluc\'{\i}a (FQM-239).


\begin{thebibliography}{99}
\bibitem{Peres93} All the proofs in ${\cal H}_3$ refers to the
spin state of a single spin-1 particle and
involve only tests of the same type, measuring
whether the spin component along two or three mutually
orthogonal unit vectors is equal to zero; see for instance,
A. Peres, Ref.\ \cite{Peres95}, p. 200.
\bibitem{remark} At this point, it is important to
mention that the multiplicative proof of the KS theorem
in ${\cal H}_4$ proposed in N. D. Mermin, Ref.\ \cite{simple},
does not exclude hidden-variable theories based on
assumptions (i) and (ii), because this proof
involves not only maximal tests containing two one-particle
observables and one two-particle observable, but also two
incompatible tests both containing two-particle observables only.
\end{thebibliography}
\end{document}